\documentclass[aps,preprint,epsfig,rotate]{revtex4}
\usepackage{graphicx}
\usepackage{bm}
\usepackage{epsfig}

\begin{document}
\title{On highly accurate computations of the Coulomb two-center systems with unit charges} 

\author{Alexei M. Frolov}
 \email[E--mail address: ]{alex1975frol@gmail.com} 


\affiliation{Department of Applied Mathematics \\
 University of Western Ontario, London, Ontario N6H 5B7, Canada}

\date{\today}

\begin{abstract}

Results of our recent highly accurate computations of the Coulomb two-center systems with the unit electrical charges $X^{+} X^{+} e^{-}$ and $X^{+} 
Y^{+} e^{-}$ are discussed. In particular, we have determined (to very high accuracy) the total energies of the ground $1 s \sigma-$states in the 
two-center adiabatic (or molecular) H$_{2}^{+}$, D$_2^{+}$, HD$^{+}$, HT$^{+}$, T$^{+}_{2}$ and DT$^{+}$ ions. In these computations we applied the 
new masses of hydrogen isotopes, which have been measured in the recent high-energy experiments. We also derived (and tested) our accurate 
mass-interpolation formula for the total energies of the model two-center $X^{+} X^{+} e^{-}$ ions with very heavy masses of the point $X^{+}$ 
particles ($M_X \ge$ 100,000 $m_e$). By using this formula we analyze the overall accuracy and validity of the adiabatic two-center approximation in 
application to the three-body atomic (or Coulomb) systems. 

\noindent 
PACS number(s): 36.10.-k and 36.10.Dr
\end{abstract}

\maketitle

\newpage
\section{Introduction}

In this study we discuss our recent results of highly accurate computations performed for the ground (bound) $1 s \sigma-$states in the two-center, 
quasi-adiabatic (or quasi-molecular) ions with the unit elecrical charges. Below we shall designate these three-body Coulomb ions by using the 
notations $X^{+} X^{+} e^{-}$ (for symmetric systems) and $X^{+} Y^{+} e^{-}$ (for non-symmetric systems). In atomic physics such two-center, 
adiabatic ions with one bound electron are called the hydrogen-like molecular ions. The label `two-center', or `adiabatic', means that the masses 
of the two heavy, positively charged particles $X$ and $Y$ are much larger than the electron mass $m_e$ (or mass of the third (negative) particle), 
i.e., $M_X \gg m_e$ and $M_Y \gg m_e$. Actual three-body adiabatic ions, which correspond to this two-center notation, are the H$_{2}^{+}$, 
D$_2^{+}$, HD$^{+}$, HT$^{+}$, T$^{+}_{2}$ and DT$^{+}$ ions. In modern high-enery physics and also in nuclear physics one finds a number of heavy 
particles $A^{+}$ which have positive (and unit) electrical charges and different masses and which are able to form a number of bound (or stable) 
two-center ions with usual neutral hydrogen (or protium), deuterium and/or tritium atoms. The newly fomed two-center ions can be deisnated as 
$(p A e)^{+}, (d A e)^{+}$ and $(t A e)^{+}$ ions, respectively. This explains an increasing interest to the two-center, one-electron systems with 
very heavy nuclear masses which are heavier and even significantly heavier, than the masses of regular hydrogen isotopes, including deuterium and 
tritium. For theoreticalpurposes we also consider the two-center Coulomb three-body ions with arbitrary nuclear masses, including very heavy and 
extremely heavy hydrogen-like nuclei with the unit electric charges. This allows us to investigate the upper mass-ratio limit and overall accuracy 
which our highly accurate exponential expansion in the relative and/or perimetric coordinates can provide for such one-electron molecular 
ions.  

Currently, there are two interesting aspects of studying adiabatic approximation for few-body systems: (a) applications to a large number of actual 
atomic (or molecular) systems, or systems with real (or realistic) atomic nuclei, and (b) improvement of the upper mass limit(s) known for our and 
other quasi-atomic variational expansions which were recently applied for highly accurate computations of the adiabatic few-body systems. In this 
study we investigate these two problems by analysing the two different groups of adiabatic (or two-center), one-electron ions. Below, for simplicity, 
we restrict ourselves to the analysis of the ground $1 s \sigma-$states in the two-center $X^{+} X^{+} e^{-}$ and $X^{+} Y^{+} e^{-}$ ions only. 
Analysis of other bound states, i.e., rotationally and vibrationally bound states, in these three-body ions can be performed analogously, but in this 
study we do not want to discuss complications which arise for those states. Note only that the ground $1 s \sigma-$states in the two-center $X^{+} 
X^{+} e^{-}$ and $X^{+} Y^{+} e^{-}$ ions have been considered in our earlier papers \cite{Fro1}, \cite{Fro2}. Analogous problems were considered in 
a number of earlier studies (see, e.g., \cite{BS} - \cite{BT}). Some of the problems discussed below were also considered in some books and textbooks 
(see e.g., \cite{Eps} and \cite{Bethe}). However, Hans Bethe decided not to include discussion of the adiabatic (or two-center), one-electron ions in 
their famous book \cite{BS} on one- and two-electron atomic systems. 

In this study we want to re-calculate the total energies and other bound state properties of the one-electron H$_{2}^{+}$, D$_2^{+}$, HD$^{+}$, 
HT$^{+}$, T$^{+}_{2}$ and DT$^{+}$ ions by using the most recent experimental masses known for these three hydrogen isotopes - protium H$^{+}$ (or 
$p$), deuterium D$^{+}$ (or $d$) and tritium T$^{+}$ (or $t$). These masses are
\begin{eqnarray}
  m_{p} = 1836.15263789 \; m_e \; \; , \; \;  m_{d} = 3670.48296785 m_e \; \; , \; \; m_t = 5496.92153588 \; m_e  \; \label{mass1} 
\end{eqnarray}
where the notation $m_e$ stands for the electron mass. These numerical values of particle masses have been determined in recent high-energy 
experiments. In particular, these masses are recomended by NIST (2019) for use in scientific research. Note also that it is very convenient to 
perform numerical calculations of all hydrogen molecular ions in atomic units and then re-calculate results of such calculations to 
electrono-Volts and/or to other units which are used in modern atomic and molecular physics. To perform such re-calculations, we need to use 
the doubled Rydberg constant which equals $2 Ry$ = 27.211386018 $eV$ and the rest mass of electron which equals $m_e$ = 0.510 998 9461 
$MeV/c^{2}$. These values of $m_e$ and $2 Ry$ are closely related to each other, since the condition $2 Ry = \alpha^{2} m_e c^{2}$, where 
$\alpha$ is the fine-structure constant $\alpha$ = 7.297352568$\cdot 10^{-3}$, is always obeyed. 

The second problem discussed in this study is the overall accuracy of adiabatic approximation. For Coulomb three-body systems the adiabatic 
approximation is often called the two-center approximation. To achieve our goal in this direction we perform highly acurate computations of 
some model adiabatic systems which have very heavy nuclear masses. Below, these ions are designated by the general notation $X^{+} X^{+} e^{-}$, 
where $M_X \ge$ 100,000 $m_e$. It is clear that we cannot find such three-body ions with unit electrical charges in everyday life. Moreover, 
even in modern high-enery experiments it is very difficult to create (and observe) very heavy nuclear particles, i.e., nuclei, or quasi-nuclei, 
with the unit electrical charges. Almost all such particles have very short life-times, and this makes practically impossible an experimental 
observation of the $X^{+} X^{+} e^{-}$ ions in which the nuclear mass of the point particle (or nuclei) $X^{+}$ exceeds 25,000 $m_e$. In this 
study we discuss such ions in order to understand (and explain) some peculiarities of the bound state energy spectra in two-center, molecular 
ions. Also, by analyzing these heavy two-center, one-electron systems we want to derive and test our original mass-asymptotic formula which is 
based on our results and can be applied to arbitrary adiabatic and symmetric $X^{+} X^{+} e^{-}$ ions.

\section{Adiabatic divergence in quantum three-body problem}

We begin our analysis from a brief discussion of the adiabatic divergence which play an important role in the general theory of quantum-mechanical 
three-body systems. The adiabatic divergence was discovered for a number of three-body variational expansions which were successfully applied 
earlier to various atomic, qausi-atomic and other `regular' Coulomb three-body systems, including the so-called `democratic' three-body systems,
where all three particle masses are comparable to each other and there is no heavy, positively charged (or central) particle. At the end of 
1980's it became finally clear that a large number of very effective and highly accurate methods developed for the both `democratic' and 
one-center Coulomb systems are practically useless in applications to the two-center Coulomb systems. In other words, these methods were found 
to be inaccurate, slow convergent and ineffective, if they are applied to the adiabatic (or two-center) Coulomb systems, such as H$^{+}_2$, 
D$^{+}_2$, HD$^{+}$, HT$^{+}$, DT$^{+}$ and T$^{+}_2$ molecular ions. Here we want to discuss this problem in detail, but first of all we need 
to investigate the principal sources of the adiabatic divergence in Coulomb three-body systems. To achieve this goal let us consider the 
Hamiltonian(s) of such systems. In this study we restrict our analysis to the Coulomb three-body systems with unit electric charges. However, 
it is clear that adiabatic divergence can also be found in other few-body systems, including Coulomb systems/ions with arbitrary electrical 
charges and systems with non-Coulomb interaction potentials. Indeed, the main source of adiabatic divergence is located in the kinetic energy of 
each three-body system (see below). Therefore, a few possible changes in the interaction potentials will not change our main conclusion. 

All systems of interest in this study, can be designated as the $(X X e)^{+}$ and $(X Y e)^{+}$ ions, where $e$ (or $e^{-}$) is the electron, 
while $X^{+}$ and $Y^{+}$ are the two heavy nuclei of hydrogen isotopes. To simplicity analysis of such systems below we discuss only the ground 
(bound) $1 s \sigma-$states in the both adiabatic $(X X e)^{+}$ and $(X Y e)^{+}$ ions. Classification of bound states in the both adiabatic 
$(X X e)^{+}$ and $(X Y e)^{+}$ ions is well explained in the literature (see, e.g., \cite{Eyr}) and here we do not want to repeat it. In the 
non-relativistic approximation the Hamiltonian $H$ of the $(X Y e)^{+}$ system in the relative coordinates $r_{32}, r_{31}$ and $r_{21}$ takes 
the following form (in atomic units)
\begin{eqnarray}
 H = - \Bigl(\frac{1}{2 m_3} + \frac{1}{2 m_2}\Bigr) \Bigl[ \frac{\partial^2}{\partial r^{2}_{32}} + \frac{2}{r_{32}} \frac{\partial}{\partial 
 r_{32}} \Bigr] - \frac{1}{2 m_3} \Bigl[\frac{r^{2}_{32} + r^{2}_{31} - r^{2}_{21}}{r_{32} r_{31}}\Bigr] \frac{\partial^2}{\partial r_{32} \partial 
 r_{31}} \nonumber \\
 - \Bigl(\frac{1}{2 m_3} + \frac{1}{2 m_1}\Bigr) \Bigl[ \frac{\partial^2}{\partial r^{2}_{31}} + \frac{2}{r_{31}} \frac{\partial}{\partial r_{31}} 
 \Bigr] - \frac{1}{2 m_2} \Bigl[\frac{r^{2}_{32} + r^{2}_{21} - r^{2}_{31}}{r_{32} r_{21}}\Bigr] \frac{\partial^2}{\partial r_{32} \partial r_{21}} 
 \nonumber \\
 - \Bigl(\frac{1}{2 m_1} + \frac{1}{2 m_2}\Bigr) \Bigl[ \frac{\partial^2}{\partial r^{2}_{21}} + \frac{2}{r_{21}} \frac{\partial}{\partial 
 r_{21}} \Bigr] - \frac{1}{2 m_1} \Bigl[\frac{r^{2}_{21} + r^{2}_{32} - r^{2}_{31}}{r_{21} r_{32}}\Bigr] \frac{\partial^2}{\partial r_{31} \partial 
 r_{21}} \nonumber \\
     - \frac{1}{r_{32}} - \frac{1}{r_{32}} + \frac{1}{r_{32}} \; \; \; \label{Hamil}
\end{eqnarray}
where the particles 1 and 2 are the two heavy $X^{+}$ and $Y^{+}$ nuclei, e.g., the nuclei of the hydrogen isotopes $p, d$ and/or $t$, while the 
light particle 3 is the electron ($e^{-}$). The three scalar relative coordinates $r_{32}, r_{31}$ and $r_{21}$ are discussed in detail in the
next Section.

As follows from Eq.(\ref{Hamil}) in the adiabatic (or two-center) limit, when $m_1 \rightarrow \infty$ and $m_2 \rightarrow \infty$ (or $\min (m_1, 
m_2) \rightarrow +\infty$), all terms which contain derivatives in respect to the $r_{21}$ variable, i.e., $\frac{\partial}{\partial r_{21}}$ and/or 
$\frac{\partial^2}{\partial r^{2}_{21}}$ derivatives, vanish from the Hamiltonian, Eq.(\ref{Hamil}). The limiting Hamiltonian $H_a$ is written in 
the form
\begin{eqnarray}
 H_a &=& - \frac{1}{2 m_3}  \Bigl[ \frac{\partial^2}{\partial r^{2}_{32}} + \frac{2}{r_{32}} \frac{\partial}{\partial r_{32}} \Bigr] - 
 \frac{1}{2 m_3} \Bigl[\frac{r^{2}_{32} + r^{2}_{31} - r^{2}_{21}}{r_{32} r_{31}}\Bigr] \frac{\partial^2}{\partial r_{32} \partial r_{31}} 
 \nonumber \\
 &-& \frac{1}{2 m_3} \Bigl[ \frac{\partial^2}{\partial r^{2}_{31}} + \frac{2}{r_{31}} \frac{\partial}{\partial r_{31}} \Bigr] - \frac{1}{r_{32}} 
 - \frac{1}{r_{32}} + \frac{1}{r_{32}} \; \; \; \label{H2+}
\end{eqnarray}
which coincides with the Hamiltonian of the model pure-adiabatic ${}^{\infty}$H$^{+}_{2}$ ion, where the both nuclei are assumed to be infinitely 
heavy, or immovable. This two-center ${}^{\infty}$H$^{+}_{2}$ ion is one-electron system has an infinite number of bound states. However, there is 
a fundamental difference between the bound state spectrum of this ion and usual energy spetra of neutral atoms and/or positively charged atomic ions. 
This difference is discissed in the Appendix A. Note that the limiting, pure adiabatic Hamiltonian $H_a$ does not contain the 
$\frac{\partial}{\partial r_{21}}$ derivative. Therefore, if $\min (m_1,m_2) \rightarrow \infty$, then the internuclear variable $r_{21} = R$ becomes 
an additional parameter of the three-body problem. In other words, the $r_{21} = R$ distance, i.e., the distance between the two heavy $X^{+}$ and 
$Y^{+}$ particles, does not change during any possible electron's motion in such a system and cannot be considered as an actual coordinate in this 
problem. In normal physical language this phenomenon is explained by the fact that the both atomic nuclei in the $(X Y e)^{+}$ system become heavier 
and heavier. The motion of these heavy nuclei begin to differ from the `regular' motion of particles in `regular' Coulomb three-body systems. For very 
heavy nuclei $X^{+}$ and $Y^{+}$ their motion is completely separated from the motion of electron. Now, we have to note that all `atomic' variational 
expansions cannot describe such a separation of nuclear and electron motions in the adiabatic (or two-center) $(X Y e)^{+}$ ions, i.e., these `atomic' 
expansions are not flexible enough. This means that we need to modify such variational expansions in order to conduct highly accurate bound state 
computations of the two-center $(X Y e)^{+}$ ions. This problem is considered in the next Section.

\section{Exponential variational expansion for the adiabatic two-center systems}

As mentioned above all highly accurate computations in this study have been performed with the use of our exponential variational expansion written 
in the relative and/or perimetric coordinates \cite{Fro1}, \cite{Fro2}. Originally, the exponential variational expansion was developed \cite{DK} 
and applied for highly accurate computations of the bound state spectra in various atomic and quasi-atomic three-body systems, such as the 
two-electron Ps$^{-}$, H$^{-}$, Li$^{+}$, Be$^{2+}$ ions, muonic $(p p \mu)^{+}$ ion and neutral He atom(s). For these and similar three-body systems 
the exponential variational expansions in the relative and/or perimetric coordinates provide very high numerical accuracy, great flexibility and 
sufficient simplicity in computations. The exponential variational expansion in the relative coordinates takes the form
\begin{eqnarray}
 \Psi(r_{32}, r_{31}, r_{21}) = \sum_{i=1}^{N} C_{i} \exp(-\alpha_{i} r_{32} - \beta_{i} r_{31} - \gamma_{i} r_{21}) \; \; \; , \; \label{equ55} 
\end{eqnarray}
where $\alpha_{i}, \beta_{i}$ and $\gamma_{i}$ (where $i = 1, \ldots, N)$ are the non-linear parameters (real numbers) of this variational expansion.
The notations $r_{32}, r_{31}, r_{21}$ in this equation designate three interparticle distances $r_{ij} = \mid {\bf r}_i - {\bf r}_j \mid = r_{ji}$ 
which are also called the relative coordinates. The wave function, Eq.(\ref{equ55}), corresponds to the bound states with $L = 0$. In fact, this 
variational expansion can be applied (with the same very high accuracy and efficiency) to the bound states with arbitrary (integer) angular momentum 
$L$ (see, e.g., discussion in \cite{Fro1}), but analysis of such cases requires a number of additional notations, and here we do not want to discuss 
the bound states with $L \ge 1$. The exponential variational expansion, Eq.(\ref{equ55}), was found to be very effective and accurate for fast 
numerical computations of various bound states in different Coulomb three-body atoms, ions and other systems. However, for adiabatic (or two-center) 
Coulomb molecular ions this variational expansion fails. For instance, in 1987 by using this expansion we could not stabilize even four decimal digits 
in the total energy of the ground $1 s \sigma-$state in the $p^{+} p^{+} e^{-}$ (or $(p p e)^{+}$) molecular ion. To solve this problem in early 
1990's we modified our exponential variational expansion, Eq.(\ref{equ55}), in the relative coordinates. The modified exponential expansion in relative 
coordinates is written in the form 
\begin{eqnarray}
 \Psi(r_{32}, r_{31}, r_{21}) = \sum_{i=1}^{N} C_{i} \exp(-\alpha_{i} r_{32} - \beta_{i} r_{31} - \gamma_{i} r_{21} - \imath \delta_{i} r_{32} 
 - \imath e_{i} r_{31} - \imath f_{i} r_{31})  \label{equ553} 
\end{eqnarray}
where $\alpha_{i}, \beta_{i}, \gamma_{i}, \delta_{i} r_{32}, e_{i}$ and $f_{i}$ are the $6 N$ non-linear parameters ($N$ is the total number of  
basis functions or exponents) of this variational expansion and $r_{32}, r_{31}$ and $r_{21}$ are the three relative coordinates. In bound state 
computations of the Coulomb adiabatic (or two-center) three-body systems our variational expansion, Eq.(\ref{equ553}), works extremely well, if the 
non-linear parameters are carefully optimized. In general, a careful optimization of these non-linear parameters in Eq.(\ref{equ553}) is an 
important step to construct highly accurate, variational wave functions and determine the bound state energies and other bound state properties of 
various adiabatic three-body systems with the unit electrical charges.  

In reality, to perform such an optimization in Eq.(\ref{equ553}) the three-body relative coordinates $r_{ij}$ are not very convenient, since they 
are not trully independent and six additional conditions $\mid r_{ik} - r_{jk} \mid \le r_{ij} \le r_{ik} + r_{jk}$, where $(i, j, k)$ = (1, 2, 3), 
must always be obeyed for three relative coordinates. These conditions drastically complicate optimization of the non-linear parameters in 
Eq.(\ref{equ553}), since instead of real optimization of the non-linear parameters one needs to check a large number of additional conditions at 
each step of the procedure. Fortunately, there is a different coordinate system for an arbitrary three-body system which is based on a different 
set of three scalar three-body coordinates which are completely independent of each other and each of them varies between 0 and $+ \infty$. These 
coordinates are the three-body perimetric coordinates $u_1, u_2$ and $u_3$ defined below. This means that by using these perimetric coordinates we 
represent the original three-dimensional space of scalar interparticle distances $r_{32}, r_{31}, r_{21}$ as a direct product of three orthogonal 
and one-dimensional spaces each of which correspond to one perimetric coordinate $u_i$ only (here $i = 1, 2, 3$). For some interaction potentials, 
e.g., for the both Coulomb and harmonic potentials, applications of these three perimetric coordinates $u_1, u_2$ and $u_3$ allows one to find even 
analytical solutions of the corresponding three-body problem(s) (see, e.g., \cite{Fro2001}). 

Such a unique combination of properties of the three-body perimetric coordinates $u_1, u_2$ and $u_3$ (independence and variation between 0 and $+ 
\infty$) make them almost perfect coordinates for 
analytical considerations and numerical computations of an arbitrary, in principle, three-body problem in Quantum and/or Classical Mechanics. Note 
that the perimetric coordnates were introduced in modern physics by C.L. Pekeris \cite{Pek1}, but, in fact, these coordinates were around for a few 
thousadns years. In particular, they were known to Hero of Alexandria, and many of us used his famous formula, which can also be written in perimetric 
coordinates \cite{Tri}, in old-fashioned (or pre-internet) schools to determine the area of a triangle using only the lengths of its edges (or ribs). 
Very likely, that Archimedes already knew and successfully used the same formula 200 years earlier. Relations between the relative and perimetric 
coordinates are simple (even linear) 
\begin{eqnarray}
  & & u_1 = \frac12 ( r_{21} + r_{31} - r_{32}) \; \; \; , \; \; \; r_{32} = u_2 + u_3 \nonumber \\
  & & u_2 = \frac12 ( r_{21} + r_{32} - r_{31}) \; \; \; , \; \; \; r_{31} = u_1 + u_3 \; \; \; \label{coord} \\
  & & u_3 = \frac12 ( r_{31} + r_{32} - r_{21}) \; \; \; , \; \; \; r_{21} = u_1 + u_2 \nonumber
\end{eqnarray}
where $r_{ij} = r_{ji}$. The Jacobian $J = \frac{\partial (r_{32}, r_{31}, r_{21})}{\partial (u_1, u_2, u_3)}$ of the $(r_{32}, r_{31}, r_{21}) 
\rightarrow (u_1, u_2, u_3)$ transformation (or substitution) of variables equals 2. As mentioned above these three perimetric coordinates $u_1, 
u_2, u_3$ are truly independent of each other and each of them varies between 0 and $+ \infty$. This drastically simplifies analytical and 
numerical computations of all three-body integrals which are needed for numerical solution of the corresponding eigenvalue problem and for 
evaluation of a large number of bound state properties in an arbitrary three-body system. The three-body exponential variational expansion in 
perimetric coordinates (for bound states with $L$ = 0) takes the form 
\begin{eqnarray}
 \Psi(u_1, u_2, u_3) = \sum_{i=1}^{N} C_{i} \exp(-\alpha_{i} u_1 - \beta_{i} u_2 - \gamma_{i} u_3 - \imath \delta_{i} u_1 - \imath e_{i} u_2 
 - \imath f_{i} u_3)  \label{equ555} 
\end{eqnarray}
where $\alpha_{i}, \beta_{i}, \gamma_{i}, \delta_{i}, e_{i}$ and $f_i$ are the $6 N-$ non-linear parameters of this expansion. These parameters are
obviously different from analogous parameters used in Eq.(\ref{equ553}). In actual applications the three last non-linear parameters (i.e., the 
$\delta_i, e_i, f_i$ parameters) in each of the basis function in Eq.(\ref{equ555}) can be chosen as arbitrary real numbers (each of them can be 
either positive, or negative, or zero), while the first three non-linear parameters $\alpha_{i}, \beta_{i}, \gamma_{i}$ must always be positive real 
numbers. It is also clear that the radial set of exponential basis functions must be a complete set. From here one finds three additional conditions 
for the $\alpha_{i}, \beta_{i}, \gamma_{i}$ parameters. In general, it can be shown that the basis of exponential `radial' functions, used in 
Eq.(\ref{equ555}), is complete, if (and only if) the three series of inverse powers of the non-linear parameters are divergent, i.e., the three 
following sums (or series): $S_1 = \sum_{i=1} \frac{1}{\alpha_{i}}, S_2 = \sum_{i=1} \frac{1}{\beta_{i}}$ and $S_3 = \sum_{i=1} \frac{1}{\gamma_{i}}$ 
are divergent when $i \rightarrow \infty$. Note that these three conditions do not complicate the actual optimization of all non-linear parameters in 
Eq.(\ref{equ555}). 

An obvious success of the exponential variational expansion, Eq.(\ref{equ555}), in applications to various three-body systems is based on very large 
numbers of carefully optimized non-linear parameters and creative strategy which was used for such an optimization. In general, there are quite a few 
different strategies which can be applied during optimization of the non-linear parameters in Eq.(\ref{equ555}). In this study we use our two-stage 
optimization strategy of the non-linear parameters in trial wave functions developed in \cite{Fro2001}. The two-stage strategy includes two equaly 
important steps: (a) very careful and accurate optimization of the non-linear parameters in the short-term wave function, and (b) optimization of the 
non-linear parameters in the rest of this trial wave function which is often called the tail. Usually, the tail of highly accurate variational wave 
function contains a few thousand terms. At the first stage we optimize each non-linear parameter in the short-term wave function \cite{Fro98}. The 
tollerance parameter for similar computations is $\approx 1 \cdot 10^{-7}$, or even smaller. Each non-linear parameter is determined (or optimized) to 
such an accuracy and then our code start to optimize another (next) non-linear parameter. When optimization of all non-linear $6 N_0$ parameters in the 
short-term wave function is finished, then the whole procedure is repeated from the very beginning. In general, optimization process for the short-term
cluster function is repeated quite a few times, e.g., until the difference between the total energies determined with this short-term wave functions and
some `exact' (or predicted) numerical value is less than $\varepsilon 1 \cdot 10^{-17}$ $a.u.$ Then we start optimization of the non-linear parameters 
in the tail part of trial wave functions is described in detail in \cite{Fro2001}. Currently, there are some alternative strategies for optimization of 
the non-linear parameters in Eq.(\ref{equ555}) which are faster and more flexible. However, it is always important to remeber that very fast and accurate 
optimization of the non-linear parameters at earlier stages of this procedure often leads to the linear dependencies of basis functions at later stages. 

\section{Highly accurate computations of the actual and realistic hydrogen molecular ions}
 
As mentioned above the experimental masses of protium, deuterium and tritium have been changed noticeably, since our previous papers were published 
in 2003 \cite{Fro2002} and even recently \cite{Fro1}, \cite{Fro2}. In general, all particle masses are always revised in numerous experiments which 
are conducted almost permanently. In this study by using the new `experimental' masses, mentioned in the Introduction, we have determined the total 
energies of the ground $1 s \sigma-$states in all hydrogen molecular ions known (and stable) in nature, i.e., in the one-electron H$_{2}^{+}$, 
D$_2^{+}$, HD$^{+}$, HT$^{+}$, T$^{+}_{2}$ and DT$^{+}$ ions. The total energies of these two-center (or molecular), one-electron H$_{2}^{+}$, 
D$_2^{+}$, HD$^{+}$, HT$^{+}$, T$^{+}_{2}$ and DT$^{+}$ ions can be found in Table I. Note that these ions play a great role in analysis of many 
problems known from stellar, atomic and molecular physics, and our new highly accurate calculations of these ions with the new nuclear masses are of 
certain interest in these areas. 

As follows from Table I the actual differences (in total energies) for the adiabatic molecular ions with the `new' and `old' nuclei of hydrogen 
isotopes used in this study (`new') and in our earlier papers (`old') \cite{Fro1}, \cite{Fro2} and \cite{Fro2002} are small. However, such 
differences are quite noticeable in modern experiments and they can lead to some deviations between experimental results and theoretical 
predictions. Therefore, our bound state computations for these ions have to be very accurate (or at least as accurate as possible). After finish of 
all such computations we could stabilize 21 - 23 decimal digits for the total energies of the ground $1 s \sigma-$states in three light H$_{2}^{+}$, 
D$_2^{+}$ and HD$^{+}$ ions. For heavier molecular ions, e.g., for the T$^{+}_{2}$ ion, we can guarantee $\approx$ 20 - 21 stable decimal digits for 
the total energies. 

Table I also shows convergencies of the total energies of the ground $1 s \sigma-$states of these ions upon the total number $N$ of basis functions (or 
exponents) used in Eq.(\ref{equ555}). Comparison of the results from Table I with analogous results obtained for these two-center H$_{2}^{+}$, D$_2^{+}$, 
HD$^{+}$, HT$^{+}$, T$^{+}_{2}$ and DT$^{+}$ ions in our earlier studies \cite{Fro1}, \cite{Fro2} allows one to evaluate the corresponding mass-gradients, 
i.e., the $\frac{\partial}{\partial M_p}, \frac{\partial}{\partial M_d}$ and $\frac{\partial}{\partial M_t}$ derivatives of the total energies for each 
of these ions. The rigorous definition of these mass gradients is
\begin{eqnarray}
 \frac{\partial}{\partial M_k} = \frac{E(M_k + \Delta_k) + E(M_k - \Delta_k) - 2 E(M_k)}{2 \Delta_k}
\end{eqnarray}
where $\Delta_k$ is the small correction to the mass $M_k$ and $k$ = 1 (or $p$), 2 (or $d$), 3 (or $t$). These mass-gradients are important values, 
since they can be used to predict changes in the total energies, if the new particle masses are applied for bound state computations of these 
two-center ions. Numerical values of these total energies are of interst in a number of applications, including astrophysycs and physics of stelar 
and laboratory plasmas. In particular, the light quasi-molecular H$_{2}^{+}$, D$_2^{+}$, HD$^{+}$ ions are of great importance for numerical 
evaluations of opacities of photospheres of some relatively cold stars from the K- and M-spectral classes, e.g., for Betelgeuse star which belongs 
to the M-class. Another area of applications of our results from Table I is numerical evaluation of the corresponding ionization thresholds for 
actual isotope-substituted hydrogen molecules. For instance, for the two-electron HD molecule its ionization threshold corresponds to the reaction 
HD = HD$^{+}$ + $e^{-}$. The energy of this threshold exactly coincides with the total energy of the quasi-molecular, one-electron HD$^{+}$ ion 
presented in Table I. This and other similar reactions of ionization of different isotope-substituted hydrogen molecules are always occur in the 
laboratory and tokamak-generated plasmas, i.e., plasmas which are needed to investigate various problems related to the nuclear fusion. 

Table II contains our highly accurate results for the total energies of the ground $1 s \sigma-$states in a number of realistic hydrogen-like 
molecular ions which are similar to the actual H$_{2}^{+}$, D$_2^{+}$ and T$_2^{+}$ ions considered above. The mass of the heavy $X^{+}$ 
particle in these realistic two-center $X^{+} X^{+} e^{-}$ ions varies between 1,000 $m_e$ and 20,000 $m_e$. Highly accurate data for these 
ions shown in Table II are needed to construct our accurate mass-interpolation formula discussed in the next Section. This mass-interpolation 
formula can be applied to approximate the total energies of the new two-center $X^{+} X^{+} e^{-}$ ions which are often observed in modern 
high-energy experiments.  

To conclude this Section let us discuss (very briefly) our results of highly accurate computations of the bound state properties of the actual 
H$_{2}^{+}$, HD$^{+}$, D$_2^{+}$, HT$^{+}$, DT$^{+}$ and T$_2^{+}$ ions. In general, if we know the wave function $\Psi$ of an arbitrary bound 
state in some quantum system, then we can determine the following expectation value   
\begin{eqnarray}
 \langle \hat{X} \rangle = \frac{\langle \Psi \mid \hat{X} \mid \Psi \rangle}{\langle \Psi \mid \Psi \rangle} = \langle \psi \mid \hat{X} \mid 
 \psi \rangle \; \; , \; \; \label{prop}
\end{eqnarray}
where $\hat{X}$ is some non-singular operator, while the wave function $\psi$ has unit norm. This definition is applied to various three-body systems, 
including Coulomb systems, i.e., atoms and ions. By choosing different operators $\hat{X}$ we can evaluate numerically a large number of bound state 
properties, e.g., electron-nuclear and inter-nuclear distances, interparticle delta-functions and other similar values. Highly accurate exponential 
variational expansions, Eq.(\ref{equ553}) and Eq.(\ref{equ555}), allow one to determine the expectation values of operators which correspond to these 
properties. Results of our numerical computations of some of the bound state properties for the ground $1 s \sigma-$states in $(p p e)^{+}, (p d e)^{+}, 
(d d e)^{+}$ and $(p t e)^{+}$ ions are shown in Table III. In this Table the indexes 1 and 2 stand for the heavy $X^{+}$ and $Y^{+}$ particles (i.e., 
for the $p, d$ and $t$ nuclei of hydrogen isotopes), while the index 3 always means the electron $e^{-}$. The properties shown in Table III include the 
$\langle r^{k}_{ij} \rangle$ expectation values, where the powers of these interparticle $(ij)$-distances are $k$ = -2, -1, 1 and 2, while $(ij)$ 
= (32) (= (31)) and (21) for the symmetric $(p p e)^{+}$ and $(d d e)^{+}$ ions and $(ij)$ = (32), (31) and (21) for the non-symmetric $(p d e)^{+}$ 
and $(p t e)^{+}$ ions. In Table III we also included the expectation values of all electron-nucleus delta-functions $\langle \delta_{31} \rangle$ and 
$\langle \delta_{32} \rangle$, the `reduced' (i.e., mass-independent) values of single particle kinetic energies $\langle - \frac12 \nabla^{2}_i 
\rangle$ of all particles and a few other properties. Since our code for such numerical computations with arbitrary arithmetic precision was not used 
for almost twenty years and we have made a number of changes in it, then we wanted to check this code in calculations of the actual hydrogen molecular 
ions with the `old' particle masses. Therefore, all our results in Table III have been obtained for the following masses of the hydrogen isotopes: 
$M_p$ = 1836.152701 $m_e$, $M_d$ = 3670.483014 $m_e$ and $M_t$ = 5496.92158 $m_e$. Coincidence the `old' and `new' results for these ions can be 
considered as excelent, which indicate clearly that our modified code for the bound state propeties is working. Note that all numerical calculations 
for this study have been performed with the use of the extended arithmetic precision (software written and later modified by David H. Bayley 
\cite{Bail1}, \cite{Bail2}).  

Physical meaning of other bound state properties shown in Table III is clear from the notation used. In particular, the notation $\tau_{ij}$ 
desigtnates the $cosine-$functions between the two corresponding inter-particle vectors, i.e.,      
\begin{eqnarray}
 \tau_{ij} = \langle \psi \mid \cos[({\bf r}_{ik})^{\wedge}({\bf r}_{jk})] \mid \psi \rangle =  \langle \psi \mid \Bigl( \frac{{\bf 
 r}_{ik}}{r_{ik}} \cdot \frac{{\bf r}_{jk}}{r_{kj}} \Bigr) \mid \psi \rangle \; \; , \; \; \label{tau}
\end{eqnarray}
where $( i, j, k)$ = (1, 2, 3). It can be shown that for an arbitrary three-body system the sum of three $cosine$ in arbitrary three-body systems is 
always represented in the form
\begin{eqnarray}
 \tau_{32} +  \tau_{31} +  \tau_{21} = 1 + 4 \langle f \rangle \; \; , \; \; \label{f}
\end{eqnarray}
where $\tau_{32} = \tau_{31}$ for symmetric systems, while the expectation value $\langle f \rangle$ is written in the form 
\begin{eqnarray}
 \langle f \rangle = \langle \psi \mid \frac{u_{1} u_{2} u_{3}}{r_{32} r_{31} r_{21}} \mid \psi \rangle = 2 \int_{0}^{+\infty} \int_{0}^{+\infty} 
 \int_{0}^{+\infty} \mid \psi(u_1, u_2, u_3) \mid^2 u_1 u_2 u_3du_1 du_2 du_3 \; \; , \; \; \label{ff}
\end{eqnarray}
The notation $\langle \delta_{3k} \rangle$ (where $k$ = 1, 2) stands for the electron-nuclear delta-functions, while the notation $\nu_{3k}$ ($k$ = 
1, 2) denotes the electron-nuclear cusps which is defined as follows 
\begin{eqnarray}
 \nu_{3k} = \frac{\langle \delta_{3k} \frac{\partial }{\partial r_{3k}} \rangle}{\langle \delta_{3k} \rangle} \; \; \; , \; \label{cusp}
\end{eqnarray}
It is the relative velocity of the two point particles (3 and $k$ in our case) at the collision point. For all Coulomb systems (and systems with 
exponential and Yukawa-type interaction potentials) the numerical values of all interparticle cusps can be determined analytically. For the 
Coulomb $X^{+} Y^{+} e^{-}$ ions such an expected (or theoretically predicted) cusp value is (see, e.g., \cite{Kato})
\begin{eqnarray}
 \nu_{3k} = q_{e} Q_{k} e^{2} \frac{m_e M_k}{m_e + M_k} = Q_{k} \frac{M_k}{M_k + 1} \; \; \; , \; \label{cuspa}
\end{eqnarray}
where $Q_k$ and $M_k$ are the electric charge (expressed in $e$) and mass (expressed in $m_e$), respectively, of the nucleus $k$. Direct comparison 
of the computed and predicted cusp values is a very good test for the quality of wave function used. As follows from Table III all electron-nuclear 
parts of our variational wave functions have a good quality. Unfortunately, the overall accuracy of these variational wave functions is very low to
determined (accurately) the nuclear-nuclear delta-function $\langle \delta_{21} \rangle$, triple delta-function $\langle \delta_{321} \rangle$ and 
nuclear-nuclear cusp value $\langle \nu_{21} \rangle$ (or cusp, for short). On the other hand, these values are of gret interest, since, e.g., the 
nuclear-nuclear delta-function $\langle \delta_{21} \rangle$ determines the corresponding fusion probabilities in each of this ion. Numerous 
attempts to evaluate these probabilities by performing highly accurate computations of these two-center (or molecular) $X^{+} Y^{+} e^{-}$ ions 
started in early 1960, but all of them were unsuccessful. 

Note that there are other bound state properties which can be computed in a few different ways. Numerical coincidence of these properties, i.e., 
the corresponding expectation values computed differently, can also be used to test the overall accuracy of our variational wave functions. In 
three-body systems one can generate a large number of usefull examples by using some basic relations between Cartesian coordinates of the three 
particles, e.g., ${\bf r}_{ij} + {\bf r}_{jk} + {\bf r}_{ki} = 0$. From here one finds $r^{2}_{ij} = r^{2}_{ik} + r^{2}_{jk} - 2 {\bf r}_{jk} 
\cdot {\bf r}_{ik}$, and therefore, for the expectation values the following equalities $\langle r^{2}_{ij} \rangle = \langle r^{2}_{ik} \rangle 
+ \langle r^{2}_{jk} \rangle - 2 \langle {\bf r}_{jk} \cdot {\bf r}_{ik} \rangle$ are always obeyed. Analogous equalities can be derived for the 
third, fourth, fifth and other higher powers $n$ of the polynomial $({\bf r}_{ij} + {\bf r}_{jk} + {\bf r}_{ki})^{n} = 0$. The same method can 
be applied to the equality which expresses the conservation of momentum in any three-body system, i.e., ${\bf p}_{1} + {\bf p}_{2} + {\bf p}_{3} 
= 0$, and therefore, $({\bf p}_{1} + {\bf p}_{2} + {\bf p}_{3})^{n} = 0$. Another approach which is often used to produce a large number of 
similar relations between different expectation values is based on the following equation which is true for an arbitrary (non-singlar) operator 
$A$ 
\begin{eqnarray}
 \langle \frac{d A}{d t} \rangle = \langle \dot{A} \rangle = \frac{\imath}{\hbar} \langle [ H, A ] \rangle = 0 \label{virial}
\end{eqnarray}
This statement is often called the Fock-Hirschfelder theorem, since Fock proved it for the virial operator $A = V = \sum^{N}_{i=1}{\bf r}_i \cdot 
{\bf p}_i$ in an arbitrary $N-$particle system (see discussion and references in \cite{Demkov}), while Hirschfelder \cite{Hir} later generalized 
this theorem for arbitrary quantum systems. By choosing different operators $A$ in Eq.(\ref{virial}) one can derive many relations $\langle [ H, A ] 
\rangle = 0$ some of which can be usefull for actual problems. Here we cannot discuss other similar equations and relations between different 
expectation values. In general, highly accurate numerical calculations of the bound state properties in the adiabatic $X^{+} X^{+} e^{-}$ and $X^{+} 
Y^{+} e^{-}$ ions is a difficult task which requires quite extensive computational resources. Indeed, the bound state computations become difficult 
to perform when the total number of basis functions used in omputations of basis functions exceeds 4,000 - 5,000 and all operations are performed 
with the computer words which contain 64 - 80 (and even 120) decimal digits. However, when such resources will be provided, then it is possible to 
finish theoretical analysis of the Coulomb adiabatic (or two-center) systems with one bound electron. 

A number of bound state properties of the symmetric $p^{+} p^{+} e^{-}$ (or H$^{+}_2$), $d^{+} d^{+} e^{-}$  (or D$^{+}_2$) and $t^{+} t^{+} e^{-}$ 
(or T$^{+}_2$) ions are presented in Table IV. In calculations of the bound state properties for Table IV we have used the nuclear masses of hydrogen 
isotopes mentioned in Introduction. Note that the bound state properties of the T$^{+}_2$ ion have never beed determined in earlier studies. Numerical
values of all bound state properties presented in Table IV are accurate, but not highly accurate. Therefore, corrections of these properties in future
works are likely. The current numerical values allow one to understand the overall geometry and basic quantum-mechanical properties of these three ions.    

\section{On accurate computations of the model hydrogen molecular ions with very heavy nuclear masses}

Thus, we have determined (to very high accuracy) the total energies of some real and realistic adiabatic H$_{2}^{+}$-like ions known in atomic 
physics. In these calculations we applied our modified exponential expansion, Eq.(\ref{equ555}), in three-body perimetric coordinates $u_1, u_2$ 
and $u_3$, which was originally developed for highly accurate computations of bound states in the one-center atomic systems and analogous 
systems/ions which have no central particle et al. Successfull applications of our exponential variational expansion, Eq.(\ref{equ555}), to various 
two-center molecular ions indicates clearly that we have developed an universal variational expansion for all non-relativistic three-body systems 
with the finite and realistic masses of particles (that goal was formulated in \cite{Fro1987}). In this Section we want to make one step beyond the 
actual mass-limits and consider what happend with our variationl expansion, Eq.(\ref{equ555}), if it is applied for bound state computations of the 
model, very heavy two-center ions with the unit electrical charges. In other words, our goal in this Section is to determine the total energies of 
the ground $1 s \sigma-$states in a number of the two-center molecular $X^{+} X^{+} e^{-}$ ions with very heavy nuclei, e.g., $M_X \ge$ 100,000 
$m_e$. This problem was considered, in part, in our earlier study \cite{Fro2}. In that parer by using the total energies computed for a number of 
model molecular ions we derived the accurate mass-interpolation (or mass-asymptotic) formula which can be applied to evaluate the total energies of 
all similar molecular $X^{+} X^{+} e^{-}$ ions, including heavy and very heavy ions. However, the actual accuracy of our mass-asymptotic formula has 
never been tested for the heavy and very heavy $X^{+} X^{+} e^{-}$ ions, since in \cite{Fro2} we could not perform direct numerical computations of 
such heavy ions. 

In this Section we determine the total energies of the ground (bound) $1 s \sigma-$states in a number of model $X^{+} X^{+} e^{-}$ ions where the 
nuclear mass $M_X$ varies between 25,000 $m_e$ and 10,000,000 $m_e$. Results of our numerical calculations (in $a.u.$) can be found in Table V. It 
was expected, that the maximal accuracy of our method will decrease, when the nuclear mass $M_X$ increases. Let us evaluate the rate at which 
our method losses its overall accuracy for heavy and very heavy adiabatic three-body $X^{+} X^{+} e^{-}$ ions. In other words, we want to find 
the threshold mass $M_X$ after which our method cannot be appled for highly accurate (and even for accurate) computations of bound states in the 
model two-center $X^{+} X^{+} e^{-}$ ions. As follows from Table V such a threshold for highly accurate computations is located somewhere around 
$M_X \approx$ 100,000 $m_e$. Actually, in current version of our computational method we do not optimize carefully all non-linear parameters. 
Futhermore, the total number of optimized non-linear parameters in our method can substantially be increased (see discussion in \cite{Fro2001}). 
These can be achieved in future computations. Very likely, by using these and a few other tricks we can increase the mass-limit of highly accurate 
computations by our method to $M_X \approx$ 300,000 $m_e$ and, probably, to larger masses. For analogous computations of less accuracy, e.g., for 
accurate calculations, or for calculations with moderate accuracy, the corresponding threshold masses are shifted to he area of larger masses (see 
Table V). 

Our results shown in Table V have different numerical accuracy, and, in general, such a level of accuracy depends upon the mass of heavy $X^{+}$ 
particle. In particular, the results obtained for different $X^{+} X^{+} e^{-}$ ions with very large masses of the  $X^{+}$ particle indicate 
clearly that our exponential variational expansion in perimetric coordinates, Eq.(\ref{equ555}), allows one to obtain accurate (but not highly 
accurate) numerical results for the adiabatic $X^{+} X^{+} e^{-}$ ions, where the mass $M_X$ varies between $\approx$ 300,000 $m_e$ and $\approx$ 
1,000,000 $m_e$. Now, it is interesting to check our interpolation formula derived in \cite{Fro2} for the total energies of some adiabatic 
two-center $X^{+} X^{+} e^{-}$ ions. As it was shown in \cite{Fro2} such a mass-interpolation formula for the adiabatic $X^{+} X^{+} e^{-}$ ions 
is based on the Puiseux series, and it can be written in the form 
\begin{eqnarray} 
 E(M) &=& \frac{M}{M + \frac12} \Bigl[ E(\infty) + \sum^{K_a}_{k=1} C_k \Bigl(\frac{1}{M}\Bigr)^{\frac{k}{q}} \Bigr] \; \; \label{Puis}
\end{eqnarray}
where $K_a$ is the maximal number of terms used in this series which can be either finite, or infinite. The parameter $q$ is the Puiseux number (or 
Puiseux parameter) which equals 4 in our case. The correct choice of this parameter is discussed in detail in \cite{Fro2} where we have noticed that 
the coefficients $C_{0}, C_{1}, \ldots, C_{k}, \ldots$ in Eq.(\ref{Puis}) take their minimal numerical values, if (and only if) the Puiseux number 
$q$ is correct. This criterion was used in \cite{Fro2} to determine the correct Puiseux number ($q$ = 4) for the adiabatic (or two-center) systems 
\cite{Fro2}. This result coindides with the Puiseux number obtained by Born and Openheimer \cite{BO} when they introduced the adiabatic (or 
Born-Openheimer) approximation. It is also clear that the first term (or $C_0$) in the right-hand side of this formula coincides with the total 
energy of the ground (bound) $1 s \sigma-$state in the true adiabatic ${}^{\infty}$H$^{+}_{2}$ ion (non-perturbed system, where $\frac{1}{M_X} = 
0$), i.e., $C_0 = E({}^{\infty}$H$^{+}_{2}) = E(\infty) \approx$ -0.60263421 44949464 6150905(5) $a.u.$ \cite{Fro2}. Numerical values of the 
following eight coefficients in this formula are (for $q$ = 4): 
\begin{eqnarray}  
  C_1 &=& -0.288279952111755252 \cdot 10^{-4} \; , \;  C_5 = -0.981049840030698796 \; , \nonumber \\
  C_2 &=&  0.227777757120415354 \; \; \; \; , \; \; C_6 = 3.77887807939899230 \; \; , \\
  C_3 &=& -0.147660324318790194 \; \; \; , \; \; C_7 = -8.53801240668919370 \; \; , \nonumber \\
  C_4 &=&  0.223497022525278858 \; \; \; \; , \; \; C_8 = 8.04015174950881455 \nonumber 
\end{eqnarray}
By using these coefficients determined in \cite{Fro2} in our asymptotic formula, Eq.(\ref{Puis}), we have evaluated the total energies of a number 
of heavy two-center $X^{+} X^{+} e^{-}$ ions, where the nuclear mass of particle $X^{+}$ equals (or exceeds) 100,000 $m_e$ \cite{Fro2}. However, at 
that time it was difficult to perform direct numerical computations of these very heavy, two-center $X^{+} X^{+} e^{-}$ ions. Since then, we could 
solve all problems with the code and obtain some additional computational resources from $sharcnet.ca$. Finally, we could finish all direct 
calculations of the heavy and very heavy two-center $X^{+} X^{+} e^{-}$ ions. Now, we can compare our earlier predictions (see Table V in 
\cite{Fro2}) with the results of direct variational computations of these ions which are presented in Table V. Such a comparison of our earlier 
predictions and direct variational results indicates clearly, that our mass-asymptotic formula is correct and accurate even for very heavy 
two-center $X^{+} X^{+} e^{-}$ ions. For instance, for the heavy two-center $X^{+} X^{+} e^{-}$ ion with $M_X$ = 50,000 $m_e$ our mass-asymptotic 
formula, Eq.(\ref{Puis}), produces $\approx$ 11 exact decimal digits for the total energy of the ground (bound) $1 s \sigma-$state. For analogous 
systems with heavier nuclear masses such a coincidence can be observed in 9 - 10 decimal digits (e.g., for the $X^{+} X^{+} e^{-}$ ions with $M_X$ 
= 100,000 $m_e$ and $M_X$ = 300,000 $m_e$). Even for extremely heavy $X^{+} X^{+} e^{-}$ ions with $M_X \ge$ 1,000,000 $m_e$ the actual agreement 
is still good (6 - 7 correct decimal digits in the total energies). Very likely, our mass-asymptotic formula, Eq.(\ref{Puis}), is substantially 
more accurate, since our results of direct variational calculations obtained for all systems are always lower then the numerical values produced 
for these systems by the formula, Eq.(\ref{Puis}). 

\section{Results and discussions}

In this study we have performed highly accurate computations of the total energies of the ground (bound) $1 s \sigma-$states for a number of adiabatic 
(or two-center), one electron $X^{+} X^{+} e^{-}$ and $X^{+} Y^{+} e^{-}$ ions. To determine the total energies of such three-body systems to very 
high accuracy we modified our exponential variational expansions in the relative and perimetric coordinates, respectively. The modified exponential 
variational expansion was found to be very effective, numerically stable and quite flexible for highly accurate computations of all real and realistic 
molecular ions. By using this expansion we have determined the total energies of the ground (bound) $1 s \sigma-$states in the two-center H$_{2}^{+}$, 
D$_2^{+}$, HD$^{+}$, HT$^{+}$, T$^{+}_{2}$ and DT$^{+}$ ions. In all such computations we applied the most recent masses of hydrogen isotopes measured 
in modern experiments. Almost all bound state properties of these ions can also be determined to high accuracy with the use of our exponential 
variational expansion in the relative and/or perimetric coordinates. However, some problems still remain in numerical computations of the 
nuclear-nuclear delta-function $\delta(r_{21})$ (or $\delta(R)$), nuclear-nuclear cusp $\nu_{21}$ and some other expectation values which include 
such a delta-function. 
 
Our computational method can also be used for highly accurate computations of the model very heavy model two-center $X^{+} X^{+} e^{-}$ and $X^{+} 
Y^{+} e^{-}$ ions. Based on this fact we investigated a number of problems known for adiabaic, two-center systems. In particular, we determine the 
actual adiabatic threshold for our exponential variational expansion, Eq.(\ref{equ555}), which was found to be equal $M = min (M_X, M_Y) \approx$ 
100,000 $m_e$. Also, we derived an accurate mass-interpolation (or mass-extrapolation, or mass-asymptotic) formula, Eq.(\ref{Puis}) for the total 
energies of the ground (bound) $1 s \sigma-$states in the adiabatic (or two-center) $X^{+} X^{+} e^{-}$ ions. This formula can be applied to the 
two-center, one-electron $X^{+} X^{+} e^{-}$ ions which are located beyond the actual (or true) adiabatic threshold known for our exponential 
variational expansion, Eq.(\ref{equ555}), e.g., to the $X^{+} X^{+} e^{-}$ ions each of which contains two positively charged particles $X^{+}$ with 
masses $M_X \ge$ 500,000 $m_e$. The total energies obtained for the ground (bound) $1 s \sigma-$states with this interpolation formula have been 
compared with the results of our direct variational computations. This comparison indicates clearly that our accurate mass-interpolation formula, 
Eq.(\ref{Puis}), provides a very good accuracy for all heavy and very heavy $X^{+} X^{+} e^{-}$ ions. 

In conclusion, we want to note that for highly accurate computations of very heavy (model) $X^{+} X^{+} e^{-}$ and $X^{+} Y^{+} e^{-}$ ions, where, 
e.g., $M = min (M_X, M_Y) \ge$ 1,000,000 $m_e$, another modification of our exponential expansion, Eq.(\ref{equ555}), is probably needed. There are 
a few possible directions for such a modification. For instance, instead of the six complex non-linear parameters our exponential variational 
expansion can include the non-linear parameters which are multiplied by quaternions. The total number of non-linear parameters equals 12 per each 
basis function, i.e., per each exponent. Such a new modification of our exponential variational expansion in three-body perimetric coordinates 
should increase the currently known abiabatic mass-limit $M_X$ of our (modified) exponential variational expansion to larger masses, e.g., to the 
$M_X \approx$ 5,000,000 $m_e$ - 10,000,000 $m_e$ values. However, we have to note again that highly accurate computations of similar, very heavy 
ions present only a very restricted, pure academic interest. Moreover, these computations can be performed more cheaply by other machinery, e.g., by 
the methods which have specifically been developed for pure adiabatic (or two-center), one-electron ions. Another interesting question which has not 
been discussed in this study is a higher symmetry of the pure adiabatic $X^{+} X^{+} e^{-}$ and $X^{+} Y^{+} e^{-}$ ions with extremely heavy $X^{+}$ 
and $Y^{+}$ nuclei. In fact, in such systems we have one additional quantum operator which commutes with the Hamiltonain of the system, i.e., this 
operator is conserved. This operator is responsible for complete separation of electron's variables in the prolate and/or oblate spheroidal coordinates 
$\xi, \eta, \phi$. In other words, when we perform a smooth transition from the actual H$^{+}_{2}$ ion to the pure adiabatic ${}^{\infty}$H$^{+}_{2}$ 
ion the actual symmetry of these three-body systems (or ions) increases. Furthermore, in addition to this actual (but relatively poor) symmetry for 
pure-adiabatic Coulomb system with one bound electron one finds a very reach `hidden' symmetry, which allows one to represent all solutions of the 
pure adiabatic (or two-center) problems as the products of two hydrogenic wave functions, i.e., two wave functions of hydrogen atoms, considered in the 
complex Cartesian space (see, e.g., \cite{Tru1}, \cite{Demk}, \cite{Fro83}). This interesting problem will be discussed in our next study. \\   

{\bf Appendix A} 

In this Appendix we want to discuss (very briefly) the general classification of bound state spectra in the Coulomb three-body systems, including 
atoms, ions and exotic systems. It is clear that the type of energy spectrum of bound states depends upon electrical charges and masses of particles. 
First, consider Coulomb three-body systems with the unit electrical charges which can be designated as $A^{+} B^{+} C^{-}$ (or $A^{-} B^{-} C^{+}$). 
Some of these systems, e.g., the $\mu^{+} \mu^{-} e^{-}$ ion, have empthy spectra of bound states. The well known Ps$^{-}$ ion has only one bound 
(ground) $1^{1}S(L = 0)-$state, while the three-body $(p p \mu)^{+}$ muonic ion has two bound states: one ground $S(L = 0)-$state and one $P(L = 
1)-$state which is the rotationally excited state. Analogous $(d d \mu)^{+}$ ion has already five bound states: two $S(L = 0)-$states, two $P(L = 
1)-$states and one $D(L = 2)-$state. The vibrationally excited (or excited) $P(L = 1)-$state in this ion is weakly-bound. For the heavier $(t t 
\mu)^{+}$ ion one finds six bound states, since one additional $F(L = 3)-$state is also stable, and none of them is weakly-bound. In the molecular 
(or two-center) $(p p e)^{+}, (d d e)^{+}$ and $(t t e)^{+}$ ions the total numbers of bound states vary between many dozens (for the $(p p e)^{+}$ 
ion) and a few hundreds (for the $(t t e)^{+}$ ion). Based on these data we can say that the trend is obvious: if the masses of heavy particles ($X$ 
and $Y$) in the both $(X^{+} X^{+} e^{-})^{+}$ and $(X^{+} Y^{+} e^{-})^{+}$ ions increase, the number of bound states in these ions also increases. 
For truly adiabatic (or two-center), very heavy $(X^{+} X^{+} e^{-})^{+}$ and $(X^{+} Y^{+} e^{-})^{+}$ ions the total number of bound states can be 
very large, e.g., a few thousands. Based on these facts, we can assume that for the two-center ions with the infinte masses of these two particles, 
i.e., for the ${}^{\infty}$H$^{+}_2$ ion, the total number of bound states must be infinite. In other words, we are dealing with the transition from 
the systems which have finite bound state spectra to the truly adiabatic three-body ${}^{\infty}$H$^{+}_2$ ion which has an infinite number of bound 
states.  

In detail, this problem was considered in our earlier paper \cite{Fro1999} where we introduced the Hamiltonian of the discrete spectrum $H_{-}$ which 
is a compact operator and has only the discrete (or bound) spectrum. The bound state spectrum of the pure-adiabatic ${}^{\infty}$H$^{+}_2$ ion is 
the spectrum of a nuclear (or kernel) compact operator $H_{-}$ which has an infinite spectrum of bound states and the sum of all its eigenvalues is 
finite (the sum of their absolute values is also finite). The general classification of bound state spectra in arbitrary Coulomb three-body (or 
$N-$body) systems is based on a number of facts known from spectral analysis of the compact operators (see, e.g, \cite{Rud}, \cite{Moren}). The first 
fundamental fact is simple: the bound state spectrum of an arbitrary $N-$body Coulomb system has no limiting points other than zero, and all non-zero 
points of this spectrum are the eigenvalues each of which has only the finite-dimensional space of eigenfunctions. Since these two conditions are 
always obeyed for an arbitrary $N-$body Coulomb system, then we have to apply the second important criterion which is based on calculation of the 
following sums:
\begin{eqnarray} 
  S_p = \sum_{i=1} \mid \lambda_i \mid^{p} = \sum_{i=1} \mid \lambda_i \mid^{p} {\rm dim}\{\phi(\lambda_i)\} = \sum_{i=1} \mid \lambda_i \mid^{p} 
 {\rm dim}\{H_{-}(\lambda_i)\} \; \; , \; \; \label{psum}
\end{eqnarray}
where $p$ is an integer parameter, $\lambda_i$ is the $i-$th discrete eigenvalue of the Hamiltonian $H_{-}$ of the bound state spectrum for this system, 
while dim$\{\phi(\lambda_i)\}$ = dim$\{H_{-}(\lambda_i)\}$ is the corresponding dimension of the $\lambda_i-$eigenspace. If the series Eq.(\ref{psum}), 
converges for $p = 0$, then the Hamiltonian $H_{-}$ of the discrete spectrum is a finite-dimensional compact operator. This is the case for all three-body 
negatively charged ions, e.g., for the Ps$^{-}$, Mu$^{-}$ and H$^{-}$ ions. Analogously, if the series, Eq.(\ref{psum}), diverges for $p = 0$, but 
converges for $p = 1$, means that the bound state spectrum of our system is generated by a nuclear compact operator $U_{-}$. In particular, this is the 
case for the pure-adiabatic ${}^{\infty}$H$^{+}_2$ ion. The third case when the series, Eq.(\ref{psum}), diverges for $p = 1$, but converges for $p = 2$, 
correspond to the Hilbert-Schmidt spectrum of some compact operator which coincides with the Hamiltonian $H_{-}$. This situation can be found in all 
neutral atoms and positively charged atomic ions, e.g., for the neutral He atom, or for the Li$^{+}$ ion(s).

\newpage
\begin{table}[tbp]
   \caption{The total energies of the ground states (or $1 s \sigma-$states) of the $(p p e)^{+}, (p d e)^{+}, (p t e)^{+}, (d d e)^{+}, 
            (d t e)^{+}$ and $(t t e)^{+}$ molecular ions in atomic units. the notation $N$ is the total number of basis functions used.
            In these calculations the `new' set of particle masses have been used.}  
     \begin{center}
     \scalebox{0.905}{%
     \begin{tabular}{| c | c | c | c |}
      \hline\hline
 $N$  & $(p p e)^{+}$ (or H$_2^{+}$) & $(d d e)^{+}$ (or D$^{+}_2$) & $(t t e)^{+}$ (or T$^{+}_2$) \\ 
     \hline   
 3000 & -0.5971390630255661737553 & -0.5987887843058796372419 & -0.5995069100987214998486 \\
 3500 & -0.5971390630255661737580 & -0.5987887843058796372501 & -0.5995069100987214998654 \\
 3800 & -0.5971390630255661737586 & -0.5987887843058796372510 & -0.5995069100987214998682 \\
 4000 & -0.5971390630255661737588 & -0.5987887843058796372515 & -0.5995069100987214998691 \\
 4200 & -0.5971390630255661737590 & -0.5987887843058796372519 & -0.5995069100987214998697 \\
 4400 & -0.5971390630255661737591 & -0.5987887843058796372520 & -0.5995069100987214998700 \\
         \hline \hline
 $N$  & $(p d e)^{+}$ (or HD$^{+})^{a}$ & $(p t e)^{+}$ (or HT$^{+}$)  & $(d t e)^{+}$ (or DT$^{+}$) \\ 
     \hline
 3000 & -0.5978979686107574611971 & -0.5981761346060211104157 & -0.5991306628357764171807 \\
 3500 & -0.5978979686107574661305 & -0.5981761346060211234832 & -0.5991306628357769125095 \\
 3800 & -0.5978979686107574669467 & -0.5981761346060211249462 & -0.5991306628357769197990 \\                         
 4000 & -0.5978979686107574673019 & -0.5981761346060211254209 & -0.5991306628357769212597 \\ 
 4200 & -0.5978979686107574675423 & -0.5981761346060211257033 & -0.5991306628357769222885 \\ 
 4400 & -0.5978979686107574676542 & -0.5981761346060211258385 & -0.5991306628357769227571 \\
         \hline \hline 
  \end{tabular}}
  \end{center}
 ${}^{(a)}$For this system the proton mass equals $m_{p}$ = 1836.15267389 $m_e$
  \end{table}
%
\begin{table}[tbp]
   \caption{The total energies of the ground states (or $1 s \sigma-$states) of some two-center, molecular ions $X^{+}X^{+}e^{-}$ 
            in atomic units. The notation $M_X$ stands for the `nuclear' mass of the heavy particle $X^{+}$ expressed in $m_e$ 
            (the mass of electron).}  
     \begin{center}
     \scalebox{0.85}{%
     \begin{tabular}{| l | l | l | l |}
      \hline\hline
 $M_X$  & $(X^{+} X^{+} e^{-})^{+}$ &  & $(X^{+} X^{+} e^{-})^{+}$ \\  
     \hline
 1000 & -0.59509329 96958491 18445  & 1500    & -0.59653116 02650119 64680 \\

 2000 & -0.59737690 59509827 42559  & 2500    & -0.59794915 10377439 18273 \\

 5000 & -0.59935177 30213303 13969  & 6000    & -0.59964364 28556994 3958 \\

 7000 & -0.59986971 95621938 2246   & 8000    & -0.60005146 72386532 8763 \\ 

 9000 & -0.60020168 15348805 3974   & 10,000  & -0.60032852 46477842 7883 \\

 11,000 & -0.6004374 799111108 5355 & 20,000  & -0.60101160 89653286 4276 \\
                  \hline\hline
  \end{tabular}}
  \end{center}
  \end{table}
\begin{table}[tbp]
   \caption{The expectation values (in atomic units) of some properties $\langle A \rangle$ for the ground  $1 s \sigma-$states in the two-center, 
   quasi-molecular $p^{+} p^{+} e^{-}, d^{+} d^{+} e^{-}, p^{+} d^{+} e^{-}$ and $p^{+} t^{+} e^{-}$ ions.} 
     \begin{center}
     \scalebox{0.95}{%
     \begin{tabular}{| l | c | c | c | c |}
        \hline\hline
 $\langle A \rangle$ & $p^{+} p^{+} e^{-}$ (or H$^{+}_2$) & $p^{+} d^{+} e^{-}$ (or HD$^{+}$) & $d^{+} d^{+} e^{-}$  (or D$^{+}_2$) & 
                       $p^{+} t^{+} e^{-}$ (or HT$^{+}$) \\
        \hline\hline        
 $\langle r^{-2}_{21} \rangle$ & 0.243 923 5001 & 0.244 842 639 & 0.245 928 3503 & 0.245 180 827 \\

 $\langle r^{-2}_{31} \rangle$ & 1.425 744 9640 & 1.429 663 740 & 1.432 641 0213 & 1.431 072 245 \\

 $\langle r^{-2}_{32} \rangle$ & 1.425 744 9640 & 1.429 817 966 & 1.432 641 0213 & 1.429 096 494 \\  
        \hline
 $\langle r^{-1}_{21} \rangle$ & 0.490 707 7799 & 0.492 957 1935 & 0.493 653 2467 & 0.492 554 1047 \\

 $\langle r^{-1}_{31} \rangle$ & 0.842 492 9625 & 0.843 714 4540 & 0.845 615 4077 & 0.844 170 9271 \\

 $\langle r^{-1}_{32} \rangle$ & 0.842 492 9625 & 0.844 138 6767 & 0.845 615 4077 & 0.844 735 4470 \\
        \hline
 $\langle r_{21} \rangle$ & 2.063 913 8668 & 2.054 803 2372 & 2.044 070 0297 & 2.051 456 6210 \\

 $\langle r_{31} \rangle$ & 1.692 966 2081 & 1.688 442 0070 & 1.680 234 6539 & 1.685 825 3635 \\

 $\langle r_{32} \rangle$ & 1.692 966 2081 & 1.687 732 4305 & 1.680 234 6539 & 1.685 825 3635 \\
        \hline
 $\langle r^{2}_{21} \rangle$ & 4.313 285 9407 & 4.268 337 208 & 4.215 643 0018 & 4.251 876 974 \\

 $\langle r^{2}_{31} \rangle$ & 3.558 797 9303 & 3.536 556 968 & 3.507 527 9324 & 3.528 357 194 \\

 $\langle r^{2}_{32} \rangle$ & 3.558 797 9303 & 3.533 836 645 & 3.507 527 9324 & 3.524 750 675 \\
        \hline 
 $\langle \frac12 p^{2}_3 \rangle$ & 0.594 292 4912 & 0.595 455 3471 & 0.596 816 0428 & 0.595 880 8216 \\

 $\langle \frac12 p^{2}_1 \rangle$ & 2.613 370 416 & 2.990 272 853 & 3.620 457 341 & 3.160 421 493 \\

 $\langle \frac12 p^{2}_2 \rangle$ & 2.613 370 416 & 2.989 147 615 & 3.620 457 341 & 3.158 860 763 \\
         \hline 
 $\langle \tau_{21} \rangle$ & 0.251 989 4930 & 0.253 062 4321 & 0.254 335 1295 & 0.253 458 2652 \\

 $\langle \tau_{31} \rangle$ & 0.509 967 7717 & 0.509 822 2948 & 0.509 171 3179 & 0.509 760 2503 \\ 

 $\langle \tau_{32} \rangle$ & 0.509 967 7717 & 0.509 384 6023 & 0.509 171 3179 & 0.590 177 8259 \\

 $\langle f \rangle$ & 0.067 981 2591 & 0.068 206 7332 & 0.068 169 4412 & 0.068 099 0854 \\
         \hline
 $\langle {\bf r}_{32} \cdot {\bf r}_{31} \rangle$ & 1.402 154 9593 & 1.401 028 2008 & 1.399 706 4302 & 1.400 615 4463 \\

 $\langle {\bf r}_{31} \cdot {\bf r}_{21} \rangle$ & 2.156 642 9711 & 2.135 528 7651 & 2.107 821 5014 & 2.127 741 7462 \\

 $\langle {\bf r}_{32} \cdot {\bf r}_{21} \rangle$ & 2.156 642 9711 & 2.132 808 4450 & 2.107 821 5014 & 2.124 135 2280 \\
         \hline 
 $\langle \delta_{31} \rangle$    &  0.206 736 53  & 0.207 043 08   &  0.207 727 14  &  0.207 160 25 \\

 $\langle \delta_{32} \rangle$    &  0.206 736 53  & 0.207 568 07   &  0.207 727 14  &  0.207 350 48 \\

 $\langle \nu_{31} \rangle$       & -0.999 457 78  & -0.999 785 18  & -0.999 730 45  & -0.999 858 61 \\

 $\nu^{(a)}_{31}$                 & -0.999 455 679 & -0.999 727 631 & -0.999 727 631 & -0.999 818 113 \\
      \hline\hline
  \end{tabular}}
  \end{center} 
  \end{table}
\begin{table}[tbp]
   \caption{The expectation values (in atomic units) of some properties $\langle A \rangle$ for the ground  $1 s \sigma-$states in the two-center, 
   quasi-molecular $p^{+} p^{+} e^{-}, d^{+} d^{+} e^{-}, p^{+} d^{+} e^{-}$ and $p^{+} t^{+} e^{-}$ ions.} 
     \begin{center}
     \scalebox{0.95}{%
     \begin{tabular}{| l | c | c | c |}
        \hline\hline
 $\langle A \rangle$ & $p^{+} p^{+} e^{-}$ (or H$^{+}_2$) & $d^{+} d^{+} e^{-}$  (or D$^{+}_2$) &  $t^{+} t^{+} e^{-}$  (or T$^{+}_2$) \\ 
        \hline\hline        
 $\langle r^{-2}_{21} \rangle$ & 0.24392 34989 & 0.24592 83520 & 0.24680 89518 \\ 

 $\langle r^{-2}_{31} \rangle$ & 1.42574 49477 & 1.43264 09459 & 1.43563 37834 \\ 
        \hline
 $\langle r^{-1}_{21} \rangle$ & 0.49070 77984 & 0.49365 32465 & 0.49494 95419 \\ 

 $\langle r^{-1}_{31} \rangle$ & 0.84249 29622 & 0.84561 54075 & 0.84698 16810 \\ 
        \hline
 $\langle r_{21} \rangle$ & 2.06391 38681 & 2.04407 00297 & 2.03538 60315 \\  

 $\langle r_{31} \rangle$ & 1.69296 62089 & 1.68234 65389 & 1.67770 76794 \\ 
        \hline
 $\langle r^{2}_{21} \rangle$ & 4.31328 59471 & 4.21564 30018 & 4.17321 44269 \\ 

 $\langle r^{2}_{31} \rangle$ & 3.55879 79326 & 3.50752 79331 & 3.48524 89025 \\ 
        \hline 
        \hline
 $\langle r^{3}_{21} \rangle$ & 9.12565 7564 & 8.77121 9415 & 8.61870 1722 \\ 

 $\langle r^{3}_{31} \rangle$ & 8.70988 1595 & 8.50374 1042 & 8.41475 2012 \\ 
        \hline 
        \hline
 $\langle r^{4}_{21} \rangle$ & 19.5423 4941 & 18.4094 6911 & 17.9279 6729 \\ 

 $\langle r^{4}_{31} \rangle$ & 24.0348 3527 & 23.2301 9151 & 22.8853 7811 \\ 
        \hline 
 $\langle \frac12 p^{2}_3 \rangle$ & 0.594 292 4910 & 0.596 816 0427 & 0.597 909 6692 \\ 

 $\langle \frac12 p^{2}_3 \rangle$ & 2.613 370 376 & 3.620 457 287 & 4.389 953 836 \\ 
         \hline 
 $\langle \tau_{21} \rangle$ & 0.251 989 49281 & 0.254 335 12958 & 0.255 371 80417 \\ 

 $\langle \tau_{31} \rangle$ & 0.509 967 77175 & 0.509 171 31777 & 0.508 819 32103 \\ 

 $\langle f \rangle$ & 0.067 9812 59076 & 0.068 1694 41277 & 0.068 2526 11554 \\ 
         \hline
 $\langle {\bf r}_{32} \cdot {\bf r}_{31} \rangle$ & 1.402 154 9592 & 1.399 706 4322 & 1.398 641 6891 \\

 $\langle {\bf r}_{31} \cdot {\bf r}_{21} \rangle$ & 2.156 642 9735 & 2.107 821 5009 & 2.086 607 2134 \\
         \hline 
 $\langle \delta_{31} \rangle$    &  0.206 736 477 & 0.207 727 552 & 0.208 151 331 \\

 $\langle \nu_{31} \rangle$       & -0.999 455 473 & -0.999 727 690 & -0.999 818 330 \\

 $\nu^{(a)}_{31}$                 & -0.999 455 679414 & -0.999 727 630495 & -0.999 818 113083 \\
      \hline\hline 
  \end{tabular}}
  \end{center} 
  \end{table}
%
\begin{table}[tbp]
   \caption{The total energies of the ground states (or $1 s \sigma-$states) in the $(X X e)^{+}$ ions (in atomic units). 
            The notation $M$ stands for the mass of the model `proton' $X$, while symbol $N$ is the total number of basis 
            functions used.}  
     \begin{center}
     \scalebox{1.05}{%
     \begin{tabular}{| c | c | c |}
      \hline\hline
 $N$  & $M$ = 25,000 $m_e$      & $M$ = 50,000 $m_e$ \\ 
     \hline
 3000 & -0.601184701103496712559 & -0.6016124409229631752 \\ 
 4000 & -0.601184701103749100204 & -0.6016124410061384276 \\ 
 4400 & -0.601184701103749100408 & -0.6016124410061384456 \\ 
      \hline 
 $A^{(a)}$ & ---------------------- & -0.601612441245840117 \\
         \hline \hline
 $N$  & $M$ = 100,000 $m_e$      & $M$ = 300,000 $m_e$ \\ 
     \hline
 3000 & -0.601913360199232050 & -0.602218387321394 \\
 4000 & -0.601913367115108117 & -0.602219110902380 \\ 
 4400 & -0.601913367115110382 & -0.602219110903757 \\
      \hline 
 $A^{(a)}$ & -0.601913366940714779 & -0.602219104988352996 \\
      \hline\hline
 $N$  & $M$ = 500,000 $m_e$      & $M$ = 1,000,000 $m_e$ \\ 
     \hline
 3000 & -0.6023101184314 & -0.602396865214 \\
 4000 & -0.6023129843035 & -0.602407341461 \\ 
 4400 & -0.6023129843254 & -0.602407343051 \\
      \hline 
 $A^{(a)}$ & -0.602312972176872087 & -0.602407318107939145 \\
      \hline\hline
 $N$  & $M$ = 5,000,000 $m_e$      & $M$ = 10,000,000 $m_e$ \\ 
     \hline
 3000 & -0.602486229 & -0.60249928 \\
 4000 & -0.602532137 & -0.60255934 \\ 
 4400 & -0.602532877 & -0.60256175 \\
      \hline 
 $A^{(a)}$ & -0.602532997346062976 & -0.602562729673433822 \\
      \hline\hline
  \end{tabular}}
  \end{center}
 ${}^{(a)}$The interpolated total energies which have been obtained from the formula, Eq.(\ref{Puis}), for these systems. 
  \end{table}

\begin{thebibliography}{99}
\bibitem{Tri} In perimeric coordinates the area of a triangle is written in the form $S_{tr} = \sqrt{u_1 u_2 u_3 (u_1 + u_2 + u_3)}$ which
explicitly shows symmetry between all vertexes of this triangle (or between all of its edges (or ribs)).
\end{thebibliography}
\end{document}